\documentclass[10pt, conference]{IEEEtran}

\usepackage{amsmath,amssymb,amsfonts}
\usepackage{algorithmic}
\usepackage{graphicx}
\usepackage{textcomp}
\usepackage{xcolor}
\usepackage{lineno}
\usepackage{float}
\usepackage{todonotes}
\usepackage{siunitx} 
\usepackage[colorlinks,allcolors=black]{hyperref}
\usepackage[doi=false,hyperref=true,datamodel=software,maxbibnames=9]{biblatex}
\usepackage[export]{adjustbox}
\usepackage{dirtree}
\usepackage{listings}

\usepackage{placeins}
\usepackage[normalem]{ulem}
\usepackage{xspace}
\usepackage{lipsum}
\usepackage{cleveref}
\usepackage{lstnix}
\usepackage{software-biblatex}

\def\RevisionsBuilt{17\xspace} %
\def\TotalDiffoscopes{\num{86476}\xspace}
\def\TotalPackagesWithDiffoscope{\num{86317}\xspace}
\def\DiffoscopeTimeout{5\xspace}
\def\HeuristicDate{15} %
\def\StartYear{2017\xspace}
\def\StartMonth{July 2017\xspace}
\def\EndYear{2023\xspace}
\def\EndMonth{April 2023\xspace}
\def\BuildHours{\num{14296}\xspace}

\begin{document}

\author{\IEEEauthorblockN{Julien Malka}
\IEEEauthorblockA{\textit{LTCI, Télécom Paris} \\
\textit{Institut Polytechnique de Paris}\\
Palaiseau, France \\
julien.malka@telecom-paris.fr}
\and
\IEEEauthorblockN{Stefano Zacchiroli}
\IEEEauthorblockA{\textit{LTCI, Télécom Paris} \\
\textit{Institut Polytechnique de Paris}\\
Palaiseau, France \\
stefano.zacchiroli@telecom-paris.fr}
\and
\IEEEauthorblockN{Théo Zimmermann}
\IEEEauthorblockA{\textit{LTCI, Télécom Paris} \\
\textit{Institut Polytechnique de Paris}\\
Palaiseau, France \\
theo.zimmermann@telecom-paris.fr}
}

\newcommand\NumberOfBuiltEvals{17} 
\newcommand\SamplingPeriod{4.1} 
\newcommand\UniquePackages{\num{58103}\xspace} 
\newcommand\TotalBuilds{\num{709816}\xspace} 
\newcommand\TotalBuildsRounded{\num{709000}\xspace} 
\newcommand\TotalBuildsReleaseNix{\num{548390}\xspace} 
\newcommand\ReleaseNixStart{\num{13527}\xspace} 
\newcommand\ReleaseNixEnd{\num{53432}\xspace} 
\newcommand\DedupDecreaseRate{21} 
\newcommand\ExpandIncreaseRate{29} 
\newcommand\PackagesIntroducedFirst{\num{13114}\xspace} 
\newcommand\PackagesIntroducedFirstStillPresent{\num{8929}\xspace} 
\newcommand\TotalEcosystems{144} 
\newcommand\PropInEcosystem{61.7} 
\newcommand\PropInTopThree{42.4} 
\newcommand\ProportionInvalid{3.2} 
\newcommand\MinNonRebuildability{0.05} 
\newcommand\MaxRebuildability{99.95} 
\newcommand\MinRebuildability{99.68} 
\newcommand\MaxNonRebuildability{0.32} 
\newcommand\TotalNonReproducibleBuilds{\num{114068}\xspace} 
\newcommand\TotalNonReproducibleBuildsRounded{\num{114000}\xspace} 
\newcommand\MinReproducibilityPropRounded{69} 
\newcommand\MaxReproducibilityPropRounded{91} 
\newcommand\ReproducibleLastRevision{90.5} 
\newcommand\BuildableLastRevision{9.4} 
\newcommand\UnreproducibilityFixedNextRevision{62} 
\newcommand\UnreproducibilityFixedNextRevisionWithRegression{85} 
\newcommand\AverageReproToBuildable{0.60} 
\newcommand\AverageBuildableToRepro{2.07} 
\newcommand\AverageGrowthRate{1.07} 
\newcommand\ReproducibleAddedPackageProp{80.02} 
\newcommand\PackagesWithEmbeddedDate{\num{12831}\xspace} 
\newcommand\PropPackagesWithEmbeddedDate{14.8} 
\newcommand\PropPackagesMatchedAtLeastOnce{19.7} 
\newcommand\PropPackagesWithEmbeddedVar{2.2} 
\newcommand\PropPackagesWithEmbeddedBuildId{2.2} 
\newcommand\PropPackagesWithEmbeddedUname{1.3} 
\newcommand\PackagesWithEmbeddedUname{\num{1097}\xspace} 
\newcommand\UnameMatchThatAreDates{\num{721}\xspace} 
\newcommand\PackagesWithDateLastRevision{\num{470}\xspace} 

\title{Does Functional Package Management\\ Enable Reproducible Builds at Scale? Yes.}

\maketitle

\begin{abstract}
  Reproducible Builds (R-B) guarantee that rebuilding a software package from source leads to bitwise identical artifacts.
  R-B is a promising approach to increase the integrity of the software supply chain, when installing open source software built by third parties.
  Unfortunately, despite success stories like high build reproducibility levels in Debian packages, uncertainty remains among field experts on the scalability of R-B to very large package repositories.

  In this work, we perform the first large-scale study of bitwise reproducibility, in the context of the Nix functional package manager, rebuilding \TotalBuilds packages from historical snapshots of the nixpkgs repository, the largest cross-ecosystem open source software distribution, sampled in the period \StartYear--\EndYear.

  We obtain very high bitwise reproducibility rates, between \MinReproducibilityPropRounded\xspace and \MaxReproducibilityPropRounded\% with an upward trend,
  and even higher rebuildability rates, over 99\%.
  We investigate unreproducibility causes, showing that about \HeuristicDate{}\% of failures are due to embedded build dates.
  We release a novel dataset with all build statuses, logs, as well as full ``diffoscopes'': recursive diffs of where unreproducible build artifacts differ.
\end{abstract}


\section{Introduction}
\label{sec:intro}

Free and open source software (FOSS) is a great asset to build trust in a computing system, because one can audit the source code of installed components to determine if their security is up to one's standards.
However trusting the source code of the components making up a system is not enough to trust the system itself: before a program can be run on a user machine, it is typically \textit{built}\footnote{A term that we will also use in this work for interpreted programs, where it is the runtime environment that has to be built.} to obtain an executable artifact, and then \textit{distributed} onto the target system, involving a set of processes and actors generally referred to as the \textit{software supply chain}.
In recent years, large scale attacks like \textit{Solarwinds}~\cite{alkhadra_solar_2021} or the \textit{xz} backdoor~\cite{noauthor_nvd_nodate} have specifically targeted the software supply chain, underlying the importance of measures to increase its security and also triggering policy response in the European Union and the United States of America~\cite{noauthor_cyber_2022, house_executive_2021}. The particular effectiveness of these attacks is due to the difficulty to analyze binary artifacts in order to understand how they might act on the system, hence increasing the need for tooling that provide traceability from executable binaries to their source code.

\textit{Reproducible builds (R-B)}---the property of being able to obtain the same, bitwise identical, artifacts from two independent builds of a software component---is recognized as a promising way to increase trust in the distribution phase of binary artifacts~\cite{lamb_reproducible_2022}. Indeed, if a software is bitwise reproducible, a user may require several independent parties to reach a consensus on the result of a compilation before downloading the built artifacts from one of them, effectively distributing the trust in these artifacts between those parties. For an attacker wanting to compromise the supply chain of that component, it is no longer sufficient to compromise only one of the involved parties.
Build reproducibility is however not easy to obtain in general, due to non-determinism in the build processes, documented both by practitioners and researchers~\cite{lamb_reproducible_2022, bajaj_unreproducible_2023}.
The \textit{Reproducible Builds}~\cite{noauthor_reproducible_2023} project has since 2015 worked to increase bitwise reproducibility throughout the FOSS ecosystems, by coming up with fixes for compilers and other toolchain components, working closely with upstream projects to integrate them.

Unfortunately, a recent study~\cite{fourne_its_2023} which interviewed 24 \mbox{R-B} experts still concluded that there is ``\textit{a perceived impracticality of fully reproducible builds due to workload, missing organizational buy-in, unhelpful communication with upstream projects, or the goal being perceived as only theoretically achievable}'' and that ``\textit{much of the industry believes [R-B] is out of reach}''.
While there exist some successful examples of package sets with high reproducibility levels like Debian, which consistently achieves a reproducibility rate of more than 95\%~\cite{noauthor_overview_nodate}, those good performances should be put in perspective with the strict quality policies applied in Debian and the relatively limited size of the package set.
Uncertainty remains among field experts about the scalability of this approach to larger software distributions.

\textit{Nixpkgs} is the largest cross-ecosystem FOSS distribution, totaling as of October 2024 about \num{100000} packages.\footnote{Based on the Repology rankings \url{https://repology.org}, accessed Oct. 2024.} It includes components from a large variety of software ecosystems, making it an interesting target to study bitwise reproducibility at scale.
Nixpkgs is built upon \textit{Nix}, the seminal implementation of the \textit{functional package management (FPM) model}~\cite{dolstra_purely_2006}.
It is generally believed that the FPM model is effective to obtain R-B: FPM packages are \textit{pure functions} (in the mathematical sense) from build- and run-time dependencies to build artifacts, described as ``build recipe''s that can be executed locally by the package manager.
Components are built in a sandboxed environment, disallowing access to unspecified dependencies, even if they are present on the system.
Previous work has highlighted that this model allows to reproduce build environments both in space and time~\cite{malka_reproducibility_2024}, a necessary property for build reproducibility.
Additionally, nixpkgs' predefined build processes implement best practices to ensure build reproducibility, like setting the \texttt{SOURCE\_DATE\_EPOCH} environment variable~\cite{noauthor_source_date_epoch_nodate} or automatically verifying that the build path does not appear in the built artifacts~\cite{noauthor_build_nodate}. Despite the potential for insightful distribution-wide build reproducibility metrics, nixpkgs limits its monitoring to the narrow set of packages included in the minimal and gnome-based ISO images~\cite{noauthor_nixos_nodate}, where a reproducibility rate higher than 95\% is consistently reported.

\paragraph*{Contributions}
In this work, we perform the first ever large scale empirical study of bitwise reproducibility of FOSS going back in time, rebuilding historical packages from evenly spaced snapshots of the nixpkgs package repository taken every \SamplingPeriod{} months from \StartYear to \EndYear. With this experiment, we answer the following research questions:
\begin{itemize}
  \item \textbf{RQ1: What is the evolution of bitwise reproducible packages in nixpkgs between \StartYear and \EndYear?} How does the reproducibility rate evolve over time? Are unreproducible packages eventually fixed? Do reproducible packages remain reproducible?
  \item \textbf{RQ2: What are the unreproducible packages?} Are they concentrated in specific ecosystems? Are critical packages more likely to be reproducible?
  \item \textbf{RQ3: Why are packages unreproducible?} Is large-scale identification of common causes possible?
   \item \textbf{RQ4: How are unreproducibilities fixed?} Are they fixed by specific patches or as part of larger package updates? Are the fixes intentional or accidental?
\end{itemize}
Besides, we use our experiment to replicate and extend previous results~\cite{malka_reproducibility_2024}, leading to an additional research question:

\begin{itemize}
\item \textbf{RQ0: Does Nix allow rebuilding past packages reliably (even if not bitwise reproducibly)?}
\end{itemize}

\paragraph*{Results}
Thanks to this large-scale experiment, we are able to establish for the first time that \textbf{bitwise reproducibility is achievable at scale}, with reproducibility rates ranging from \MinReproducibilityPropRounded{}\% to \MaxReproducibilityPropRounded{}\% over the period \StartYear--\EndYear,
despite a continuous increase in the number of packages in nixpkgs.
  We highlight the wide variability in reproducibility rates across ecosystems packaged in nixpkgs, and show the significant impact that some core packages can have on the overall reproducibility rate of an ecosystem.

  We estimate the prevalence of some common causes of non-reproducibility at a large scale for the first time, showing that about \textbf{\HeuristicDate\% of failures are due to embedded build dates}.

  As part of this work, we introduce a \textbf{novel dataset containing build logs and metadata} of over \TotalBuildsRounded package builds, and more than \TotalNonReproducibleBuildsRounded occurrences of non-reproducibility with full artifacts including ``diffoscopes'', i.e., recursive diffs of where unreproducible build artifacts differ.
  Ample room for further research is left open by the dataset, including exploiting the build logs, or applying more complex heuristics or qualitative research to the diffoscopes.

\paragraph*{Paper structure}%
\Cref{sec:related} presents the related work. \Cref{sec:context} gives some background that is required to understand the experiment, whose methodology is then presented in \Cref{sec:methodology}. Some descriptive statistics about the dataset are presented in \Cref{sec:data}, and the results to our RQs in \Cref{sec:results}. We discuss them in \Cref{sec:discussion}, and the threats to validity in \Cref{sec:threats}, concluding in \Cref{sec:conclusion}.

\section{Related work}
\label{sec:related}

\subsection{Reproducible builds (R-B)}

R-B are a relatively recent concept, which has been picked up and developed mostly by practitioners from Linux distributions and upstream maintainers. The \textit{Reproducible Builds} project~\cite{noauthor_reproducible_2023} has been the main actor in the area. %
The project has produced a definition of R-B, best practices to achieve them, and tools to monitor the reproducibility of software distributions, and debug unreproducibilities (the \textit{diffoscope}).

Besides, R-B have picked the interest of the academic community, with a growing number of papers on the topic.

\paragraph{R-B for the security of the software supply chain}

R-B are often seen as a way to increase the security of the software supply chain~\cite{lamb_reproducible_2022}. Torres-Arias \emph{et al.} provide a framework to enforce the integrity of the software supply chain for which they demonstrate an application to enforce R-B~\cite{torres-arias_-toto_2019}.
Our paper does not contribute directly to this line of research but, by demonstrating the feasibility of R-B at scale, it strengthens the case of the approach.

\paragraph{Techniques for build reproducibility}

In a series of articles~\cite{ren_automated_2018,ren_root_2020,ren_automated_2022}, Ren \emph{et al.} devised a methodology to automate the localization of sources of non-reproducibility in build processes and to automatically fix them, using a database of common patches that are then automatically adapted and applied.

An alternative technique to achieve build reproducibility is proposed by Navarro \emph{et al.}~\cite{navarro_leija_reproducible_2020}. They propose ``reproducible containers'' that are built in a way that makes the build process fully deterministic, at the expense of performance.

FPMs such as Nix~\cite{dolstra_purely_2006} and Guix~\cite{courtes_functional_2013} are also presented as a way to achieve R-B. %
Malka \emph{et al.}~\cite{malka_reproducibility_2024} showed that Nix allows reproducing past build environment reliably, as well as rebuilding old packages with high confidence, but they do not address the question of bitwise reproducibility, which we do with this work.

\paragraph{Relaxing the bitwise reproducibility criterion}

Because of the difficulty (real or perceived) to achieve bitwise reproducibility, some authors have proposed to relax the criterion to a more practical one. For instance, \emph{accountable builds}~\cite{poll_analyzing_2021} aim to distinguish between differences that can be explained (accountable differences) or not (unaccountable differences).
Our work highlights that bitwise reproducibility is achievable at scale in practice, and thus that relaxing the reproducibility criterion may not be necessary after all.

\paragraph{Empirical studies of R-B}

Some other recent academic works have empirically studied R-B in the wild. Two papers from 2023~\cite{butler_business_2023,fourne_its_2023} looked into business adoption of R-B and perceived effectiveness through interviews.

Bajaj \emph{et al.}~\cite{bajaj_unreproducible_2023} mined historical results from R-B tracking in Debian to investigate causes, fix time, and other properties of unreproducibility in the distribution. Our work is similar, but instead of relying on historical R-B tracking, we actually rebuild packages and compare them bitwise to historical build results.
When we report on packages being reproducible, it means they have stood the test of time.
It also allows us to provide more detailed information on the causes of unreproducibility, in particular by generating diffoscopes and saving them for future research as part of our dataset; whereas diffoscopes from the Debian R-B tracking are not preserved in the long-term.
Finally, Bajaj \emph{et al.} used issue tracker data from the R-B project to identify the most common causes of non-reproducibility, possibly introducing a sampling bias since only root causes that were identified by Debian developers are counted in their statistics. In our work, we try to avoid this bias by performing a large-scale automatic analysis of diffoscopes to automatically identify the prevalence of a selection of causes of non-reproducibility.
While we present heuristics comparable to some of the causes identified in Bajaj \emph{et al.}'s taxonomy, we derive them from empirical data rather than relying on pre-labeled data from the Debian issue tracker.

The only other work that performed an experimental study of R-B investigated the impact of configuration options~\cite{randrianaina_options_2024}. Contrary to them, we rebuild historical versions of packages in their default configuration. Combining the historical snapshot approach of our work with their approach of varying the configuration options could be an interesting future work.

\subsection{Linux distributions and package ecosystems}

Besides R-B, our work also relates to the literature on Linux distributions and package ecosystems. The nixpkgs repository being the largest cross-ecosystem software distribution, we are able to compare properties of packages across ecosystems. Several previous works have compared package ecosystems (\emph{e.g.,} \cite{decan_empirical_2019}). For an overview of recent research on package ecosystems, see Mens and Decan~\cite{mens_overview_2024}.

More specifically, nixpkgs is the basis of the NixOS Linux distribution. Linux distributions have a long history of being studied by the research community. Recently, Legay \emph{et al.}~\cite{legay_quantitative_2021} measured the package freshness in Linux distributions. While this is not the topic of this work, our dataset could be used, e.g., to study how frequently packages are updated in nixpkgs.

\section{Background}
\label{sec:context}

We provide in this section some background knowledge about Nix and R-B, which is required to understand the details of our experiments.

\subsection{The FPM model and the Nix store}

The FPM model applies the idea of functional programming to package management. Nix packages are viewed as pure functions from their inputs (source code, dependencies, build scripts) to their outputs (binaries, documentation, etc.).
Any change to the inputs should produce a different package version.
Nix allows multiple versions of the same package to be built and coexist on the same system.
To that end, Nix stores build outputs in input-addressed directories (using a hashing function of the inputs) in the \textit{Nix store}, usually located in the \texttt{/nix/store} directory on disk.
\Cref{fig:nix_expression} shows an example of a Nix packaging expression (a \emph{Nix recipe}) for the \texttt{htop} package.
\begin{figure}
\begin{adjustbox}{max width=\linewidth}
\begin{lstlisting}[language=Nix, escapechar=|]
{stdenv, fetchFromGitHub, ncurses, autoreconfHook}:

stdenv.mkDerivation rec {
  pname = "htop";
  version = "3.2.1";

  src = fetchFromGitHub {
    owner = "htop-dev";
    repo = "htop";
    rev = version;
    sha256 = "sha256-MwtsvdPHcUdegsYj9NGyded5XJQxXri1IM1j4gef1Xk=";
  };

  nativeBuildInputs = [ autoreconfHook ];
  buildInputs = [ ncurses ];
  };
}

\end{lstlisting}
\end{adjustbox}
\caption{Example Nix expression for the \texttt{htop} package.}
\label{fig:nix_expression}
\end{figure}

\subsection{Nix evaluation-build pipeline}

Building binary outputs from a Nix package recipe is a two-step process.
First, Nix \textit{evaluates} the expression and transforms it into a \textit{derivation}, an intermediary representation in the Nix store containing all the necessary information to run the build process. In particular, the derivation contains ahead of time the (input-addressed) \textit{output path}, that is the exact location in the Nix store where the build artifacts will be stored if that derivation were to be built.
Then, given the derivation file as input, the \texttt{nix-build} command performs the build, creating the pre-computed output path in the Nix store upon completion.

In the same fashion as other Linux distributions, Nix packages may produce multiple outputs (a main output with binaries, one with documentation, etc.).
Each output has its own directory in the Nix store, and building the derivation from source systematically produces all its outputs.

\subsection{Path substitution and binary caches}

Alternatively to building from source, Nix offers the option to download prebuilt artifacts from third party \emph{binary caches}, which are databases populated with build outputs generated by Nix.
Binary caches are indexed by output paths, making it possible for Nix to check for the presence of a precompiled package in a configured cache after the evaluation phase.
\url{https://cache.nixos.org} is the official cache for the Nix community and most Nix installations come configured to use it as a trusted cache.

\subsection{The nixpkgs continuous integration}

Hydra~\cite{dolstra_nix_nodate} is the continuous integration (CI) platform for the nixpkgs project.
At regular intervals in time, it fetches the latest version of nixpkgs' git \texttt{master} branch and evaluates the \texttt{pkgs/top-level/release.nix} file embedded in the repository.
This evaluation yields a list of derivations (or \textit{jobs}) that are then built by Hydra: one derivation for each of the $\approx\,$\num{100 000} packages contained in nixpkgs nowadays.
Upon success of a predefined subset of these jobs, the revision is deemed valid and all the built artifacts are uploaded to the official binary cache to be available to end users.

\subsection{Testing bitwise reproducibility with Nix}
\label{sec:check}

Nix embarks some minimal tooling to test the reproducibility of a given derivation in the form of a \texttt{--check} flag passed to the \texttt{nix-build} command. To check for bitwise reproducibility, Nix needs a reference that it will try to acquire from one of the configured caches, or fail if not possible. Nix then acquires the build environment of the derivation under consideration, builds the derivation, and compares each of the outputs of the derivation against the local version. The \texttt{--keep-failed} flag can be used to instruct Nix to keep the unreproducible outputs locally for further processing.\footnote{For our experiment, we alter the behavior of \texttt{nix-build --check} to prevent it from failing early as soon as one unreproducible output is detected.}

\subsection{Diffoscope}

Diffoscope~\cite{noauthor_diffoscope_nodate} is a tool developed and maintained by the Reproducible Builds project that aims to simplify the analysis of differences between software artifacts. It is able to recursively unpack binary archives and automatically use ecosystem specific diffing tools to allow for better understanding of what makes two software artifacts different. It generates HTML or JSON artifacts---also called \textit{diffoscopes}---that can be either interpreted by humans or automatically processed.

\section{Methodology}
\label{sec:methodology}

\begin{figure*}
  \centering
  \includegraphics[width=0.95\textwidth]{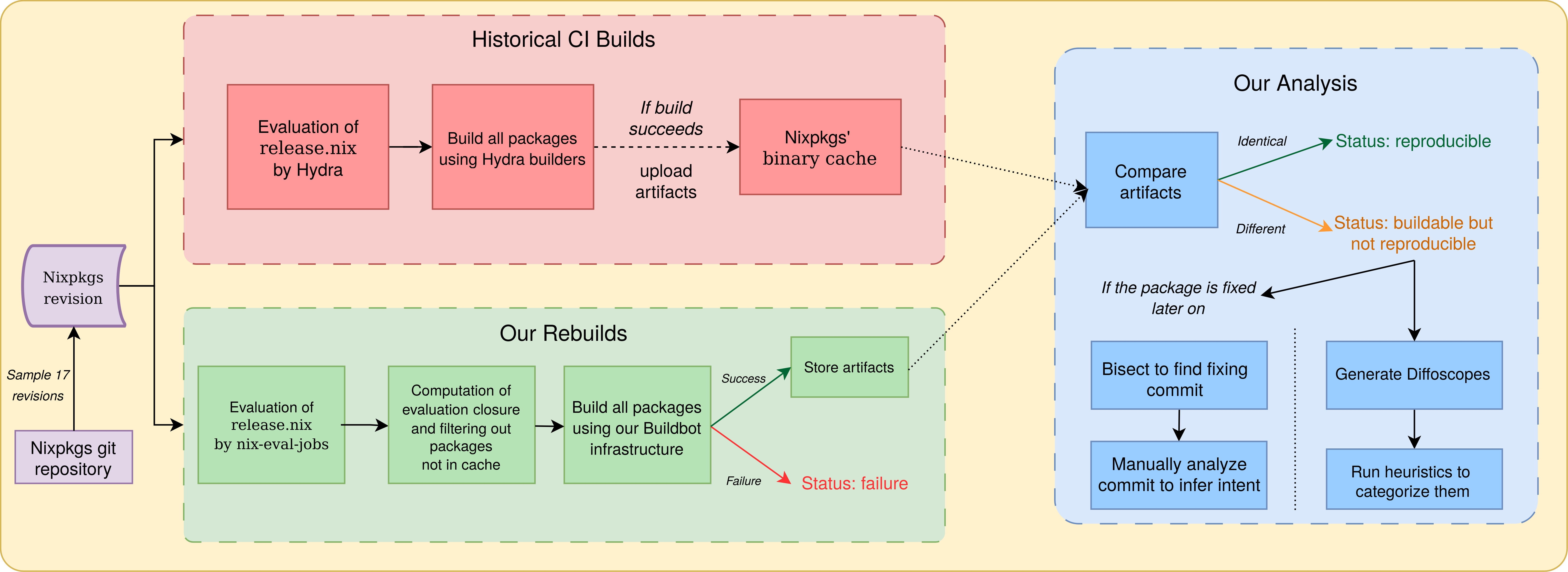}
  \caption{Description of our build and analysis pipeline.}
  \label{fig:experimental-pipeline}
\end{figure*}

Our build and analysis pipeline is summarized in \Cref{fig:experimental-pipeline}.

\subsection{Reproducibility experiments}

\subsubsection{Revision sampling}

We start from the 200 nixpkgs revisions selected by Malka \emph{et al.}~\cite{malka_reproducibility_2024} in 
the period \StartMonth--\EndMonth.
Since \emph{building} revisions, as opposed to just evaluating them, is very computationally intensive, it was not feasible to build all 200 revisions.
Also, it was difficult to correctly estimate how many revisions we could build, due to the ever-growing number of packages in each revision.
We hence applied dichotomic sampling: we first build the most recent revision, then the oldest one, then the one in the middle of them, and so on always picking the one in the middle of the largest time interval when choosing.
After \RevisionsBuilt{} revisions built, we obtain a regularly spaced sample set of nixpkgs revisions, with one sampled revision every \SamplingPeriod{} months.
On average, each revision corresponds to building more than 41 thousand packages (see \Cref{fig:packageset-evolution} for details, discussed later).

To perform our builds, we used a distributed infrastructure based on Buildbot~\cite{noauthor_buildbot_nodate}, a Python CI pipeline framework. Our infrastructure has two types of machines: a \textit{coordinator} and multiple \textit{builders}. The coordinator is in charge of distributing the workload and storing data that must be persisted, while the builders are stateless and perform workloads sent by the coordinator. During the course of the experiment (from June to October 2024) the set of builders we used was composed of shared bare-metal machines running various versions of Ubuntu and Fedora and our coordinator was a virtual machine running Ubuntu.
Note that to perform our bitwise reproducibility checks, we compare to the historical results coming from Hydra, that uses builders that, at the time, ran older versions of Nix than the ones we used on our builders.\footnote{Further details on the operating systems, Nix versions and kernel versions that we used on builders can be found in the replication package.}

\subsubsection{Evaluation and preprocessing}

For each revision considered, the coordinator first evaluates \texttt{pkgs/top-level/release.nix} (containing the list of jobs built by Hydra for this revision) using \texttt{nix-eval-jobs}, a standalone and parallel Nix evaluator similar to the one used on Hydra.
The outcome of this operation is a list of derivations.
The \texttt{release.nix} file is human crafted and does not contain all the dependencies of the listed packages, even though they are built by Hydra along the way.
Since we are interested in testing the reproducibility of the entire package graph built by Hydra, we post-process the list of jobs obtained after the evaluation phase to include all intermediary derivations by walking through the dependency graph of each derivation.
During this post-processing, we also check that the derivation outputs are present in the official Nix binary cache.
This is required to compare our build outputs with historical results for bitwise reproducibility.
Derivations missing from the cache can indicate that they historically failed to build, although there can be other reasons for their absence.

\subsubsection{Building}

To build a job $A$ and test the reproducibility of its build outputs, the builder uses the \texttt{nix-build --check} command, as described in \Cref{sec:check}.
This means that we always assume that $A$'s build environment is buildable and always fetch it from the cache. This allows all the derivations in the sample set to be built and checked independently and in parallel, irrespective of where they are located in the package dependency graph.
Note that the source code of packages to build is part of the build environment.
Relying on the NixOS cache hence avoids incurring into issues such as source code disappearing from the original upstream distribution place.
Investigating how much of NixOS can be rebuilt without relying on the cache is an interesting research question, recently explored for Guix~\cite{courtes-2024-guix-archiving}, but out of scope for this paper.

After each build, we classify the derivation as either \textit{building reproducibly}, \textit{building but not reproducibly} or \textit{not building}.
We save the build metadata and the logs, and when available we download and store the historical build logs from the nixpkgs binary cache. Finally, for every unreproducible output path, we store both the historical artifacts and our locally built ones, for comparison purposes.

\subsection{Ecosystem identification and package tracking}

To answer RQ1, RQ2 and RQ4, we need to be able to discriminate packages by provenance ecosystem, and track them over time to follow their evolution.
To categorize packages by ecosystem, we rely on the first component of the package name when it has several components (for example a package named \texttt{haskellPackages.network} is sorted into the Haskell ecosystem). Sometimes, there are several co-existing versions of an ecosystem in a given nixpkgs revision (for example \texttt{python37Packages} and \texttt{python38Packages} being present in the same revision), and sometimes the name of the ecosystem is modified between successive nixpkgs revisions. Therefore, some deduplication step is necessary. The first and last authors performed this step manually by inspecting the \TotalEcosystems{} ecosystems from the \RevisionsBuilt{} nixpkgs revisions considered, ordered alphabetically, and deciding which ones to merge independently, then checking the consistency of their results, discussing the few differences (missed merges, or false positives) and reaching a consensus. For instance, the following ecosystems were merged into a single one: \texttt{php56Packages}, \texttt{php70Packages}, \ldots, \texttt{php82Packages}, \texttt{phpExtensions}, \texttt{phpPackages} and \texttt{phpPackages-unit}.

To deduplicate packages appearing in several versions of the same ecosystem, we order by version (favoring the most recent one) and consider any package set without a version number as having a higher priority (since it is the default one in the considered revision, as chosen by the nixpkgs maintainers).

\subsection{Comparison with the minimal ISO image}

As part of RQ2, we investigate the difference of reproducibility rate between critical packages whose reproducibility is monitored and the rest of the package set. We are also interested in knowing whether observing the reproducibility health of this subset of packages gives a good enough information on the state of the rest of the project. The minimal and gnome-based ISO images are considered critical subsets of packages and benefit from a community-maintained reproducibility monitoring. We study the minimal ISO image because it contains a limited amount of core packages. We evaluate the Nix expression associated with the image, compute its runtime closure (the set of packages included in the image) and match it with the packages of our dataset to infer their reproducibility statuses.

\subsection{Analyzing causes of unreproducibility using diffoscopes}

Analyzing causes of unreproducibility is a tricky debugging activity, usually carried out by practitioners (in particular, by Linux distribution maintainers and members of the Reproducible Builds project).
Some automatic fault localization methods have been proposed~\cite{ren_root_2020}, but they rely on instrumenting the build, while we have to run the Nix builds unchanged to avoid introducing biases.

For each unreproducible output, we run diffoscope with a \DiffoscopeTimeout-minute timeout, yielding a dataset of \TotalDiffoscopes diffoscopes. We then investigate whether we can use our large dataset of diffoscopes for automatic detection of causes of non-reproducibility. The diffoscope tool was mainly designed to help human debugging, but it also supports producing a JSON output, which can then be machine processed.

We wish to explore heuristics that can be applied at the line level, so we recurse through diffoscope structures until leaf nodes, which are diffs in unified diff format.
We randomly draw one added line from \num{10000} diffoscopes, sort them by similarity to ease visual inspection, and manually inspect them to derive relevant heuristics.
We then run these heuristics on the full diffoscope dataset to determine the proportion of packages impacted by each cause (multiple causes can apply to the same package).
The first and last author then evaluate the precision of each these heuristics by manually counting false positives in samples of matched lines for each heuristic.

\subsection{Automatic identification of reproducibility fixes}

To investigate fixes to unreproducibilities, for each unreproducible package that becomes reproducible, we run an automatic bisection process to find the first commit that fixes the reproducibility. By looking into the corresponding pull request on GitHub, we check if the maintainers provide information on why this fixes a reproducibility issue, or link to a corresponding bug report. In particular, we are interested to check how often the maintainers are aware that their commit is a reproducibility fix (as opposed to a routine package update, which embeds a reproducibility fix that would have been crafted by the upstream maintainers).

We start from the set of packages from all revisions, and we look specifically at packages that change status from unreproducible to reproducible in two successive revisions.
These are our candidate packages (and ``old'' and ``new'' commits) for the bisection process.

Since we perform the bisection process on a different runner and at a different time compared to the dataset creation, it can happen that we cannot reproduce the status (reproducible or unreproducible) of some builds. Therefore, before starting the bisection, we verify that we obtain consistent results on the ``old'' and the ``new'' revision. %
Then, for those which behave as expected, we start an automatic \texttt{git bisect} process.

The script used for the automatic \texttt{git bisect} checks for the reproducibility status of the build to mark the selected commit as ``old'' or ``new''.
Commits that fail to build or are not available in cache are marked as ``skipped''. %
We use \texttt{git bisect} with the \texttt{--first-parent} flag because intermediate pull request commits are typically not in cache.

For the qualitative analysis, we first group packages by fixing commit, as seeing all packages fixed by a given commit gives valuable information that might help to understand the reproducibility failure being fixed.
We then randomly sample fixes and open their commit page on GitHub, locating the corresponding pull request.
We manually inspect the pull request (description, commit log, code changes) to first confirm that the bisect phase successfully identified the commit that fixed the reproducibility failure.
It may be the case that after careful inspection, the change looks unrelated to the package being fixed (the bisection process can give incoherent results in case of a flaky reproducibility issue or because the package changed status several times between two data points) in which case we discard it.
Once we have confirmed that the identified commit is correct, we check whether the commit authors indicate that they are fixing a reproducibility issue and if the commit is a package update or another change.
We analyze 100 randomly sampled reproducibility fixes and report our findings.
Additionally, we perform the same analysis on the 15 commits that fix the most packages (from \num{3052} down to 27 packages fixed) to find potential differences of behavior of the contributors for those larger-scale fixes.

\section{Dataset}
\label{sec:data}

The main result of running the pipeline of \Cref{fig:experimental-pipeline} is a large-scale dataset of historical package rebuilds, including (re)build information, bitwise reproducibility status and, in case of non-reproducibility, generated diffoscopes.
In this paper, we use the dataset to answer our stated research questions, but many other research questions could be addressed using the dataset, including more in-depth analysis of non-reproducibility causes.
We make the dataset available to the research and technical community to foster further exploration on the topic.
In the remainder of this section, we provide some descriptive statistics of the dataset.

\begin{figure}
  \includegraphics[width=0.47\textwidth, right]{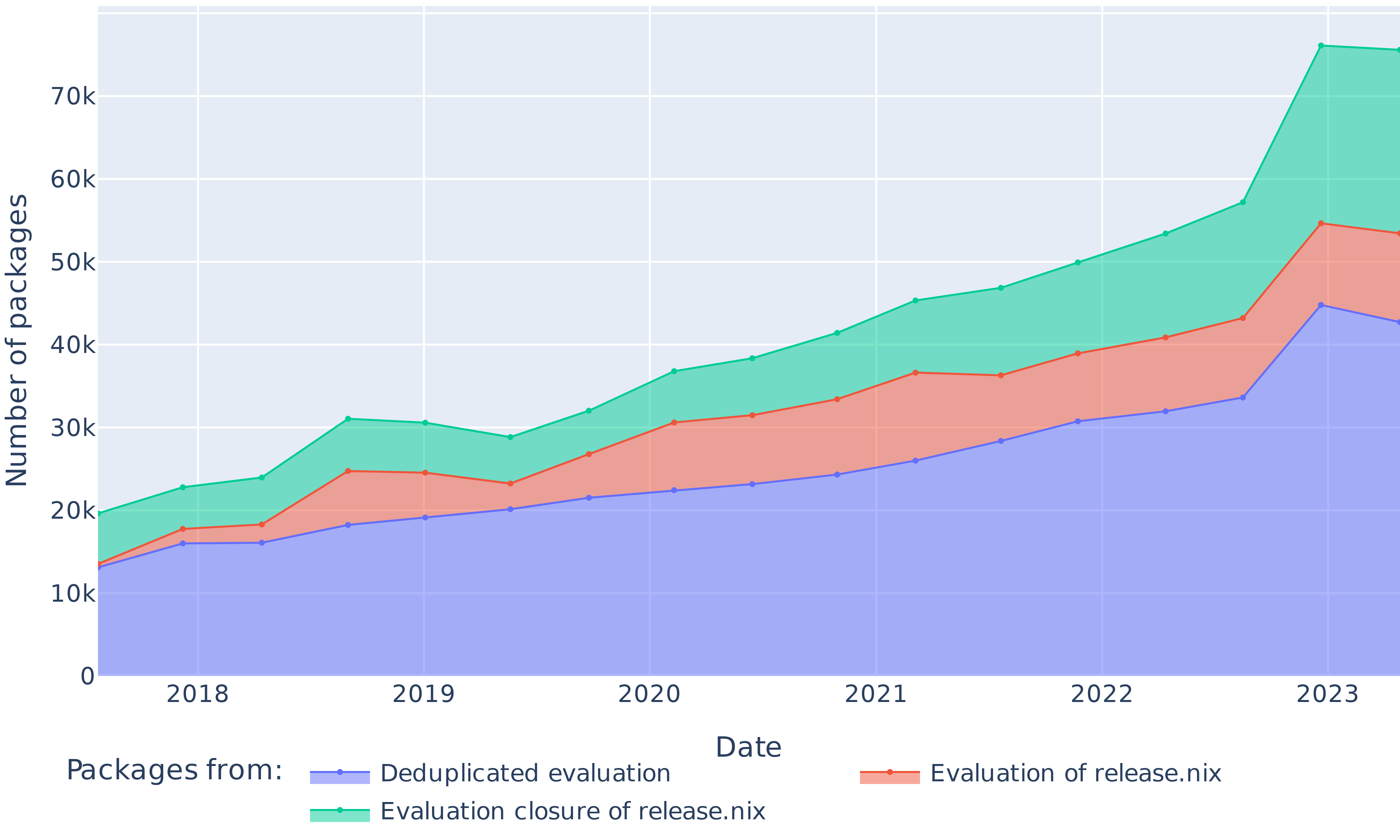}
  \caption{Evolution of the number of packages in each nixpkgs revision (as defined by \texttt{release.nix}), in their evaluation closure and after deduplicating ecosystem copies.}
  \label{fig:packageset-evolution}
\end{figure}

The dataset spans \TotalBuilds package builds coming from \RevisionsBuilt{} nixpkgs revisions, built over a total of \BuildHours hours.
From those builds, \TotalBuildsReleaseNix are coming directly from the \texttt{release.nix} file and can be tracked by name.
They correspond to \UniquePackages unique packages that appear over the span of sampled revisions.
As can be seen on \Cref{fig:packageset-evolution}, the number of packages listed to be built by Hydra increased from \ReleaseNixStart in \StartYear to \ReleaseNixEnd in \EndYear.
Ecosystems can be present in multiple versions.
On average, deduplicating packages in multiple ecosystem copies decreases their number by \DedupDecreaseRate{}\% while adding the evaluation closure of the \texttt{release.nix} file increases the number of jobs by \ExpandIncreaseRate{}\%.

\begin{figure}
  \includegraphics[width=0.47\textwidth, right]{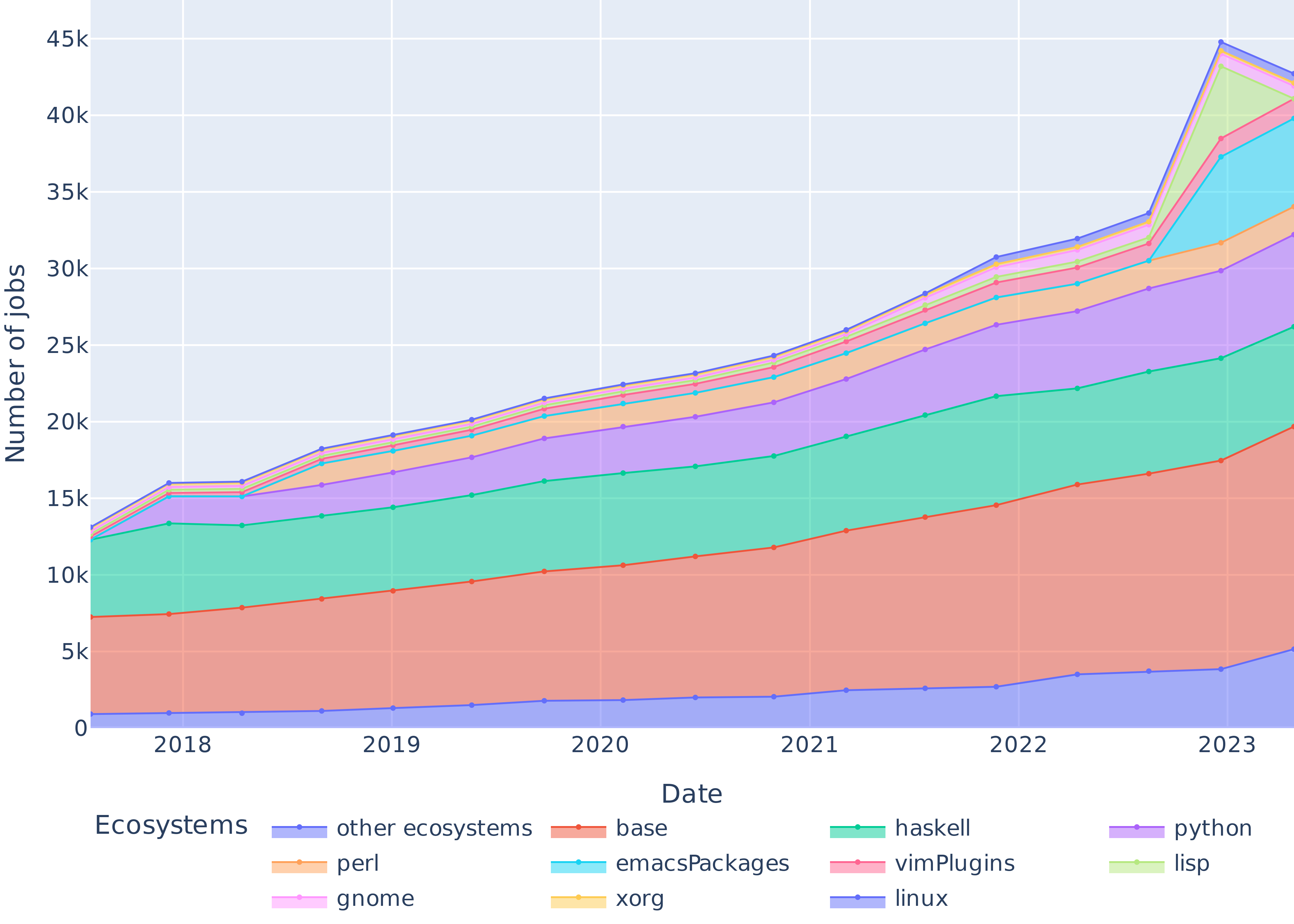}
  \caption{Evolution of the size of the nine most popular software ecosystems in nixpkgs, the packages whose ecosystem is undetermined (base), and the packages from other ecosystems (only packages listed in \texttt{release.nix}).}
  \label{fig:evolution-ecosystems}
\end{figure}

\Cref{fig:evolution-ecosystems} shows the evolution of the top 9 ecosystem sizes in nixpkgs, plus the base namespace.
\PropInEcosystem{}\% of packages belong to an ecosystem, while the rest live in nixpkgs base namespace.
The three largest ecosystems are Haskell, Python and Perl, which together account for \PropInTopThree{}\% of the packages.

\begin{figure}
  \includegraphics[width=0.47\textwidth, right]{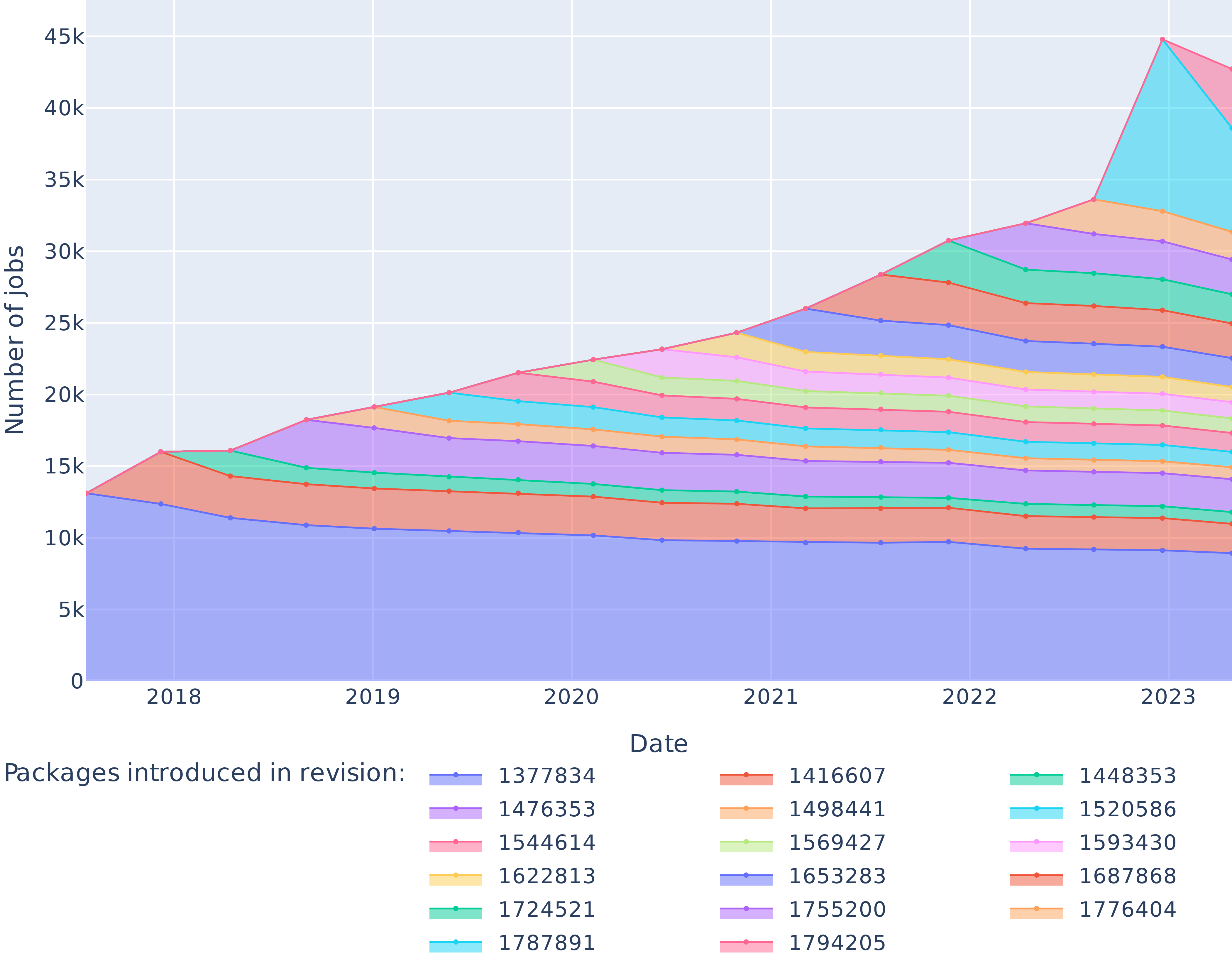}
  \caption{Number of packages introduced by every revision of the dataset and their survival in the package set over time.}
  \label{fig:survival-packages}
\end{figure}

\Cref{fig:survival-packages} outlines the number of packages introduced in the dataset by each revision, and their survival over time.
In particular, as of \StartMonth the package set contained \PackagesIntroducedFirst elements, \PackagesIntroducedFirstStillPresent of which were still present in \EndMonth.

\section{Results}
\label{sec:results}

We present our experimental results below, organized by research question.
Their discussion is provided later, in \Cref{sec:discussion}.

\subsection{RQ0: Does Nix allow rebuilding past packages reliably (even if not bitwise reproducibly)?}

\begin{figure}
  \includegraphics[width=0.47\textwidth, right]{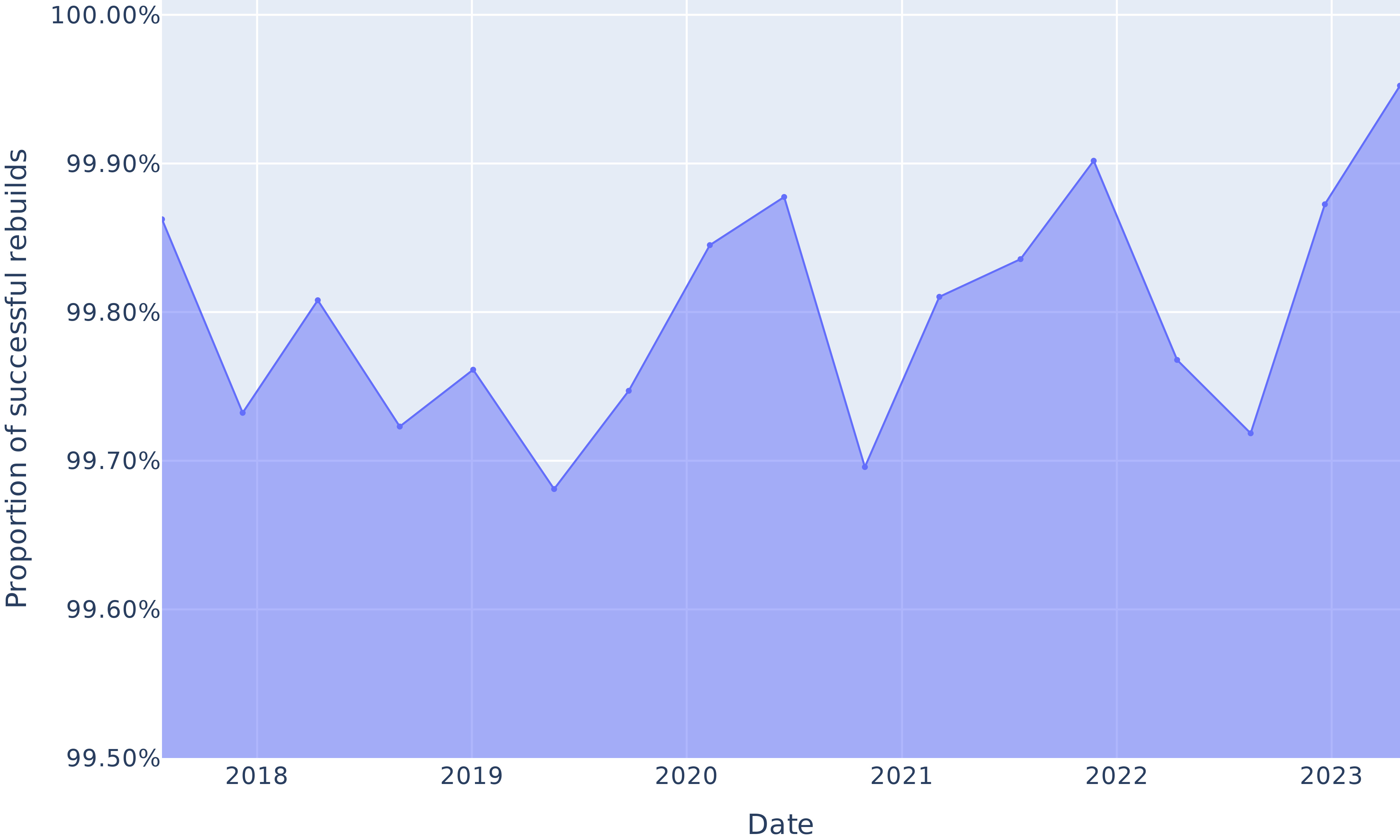}
  \caption{Proportion of rebuildable packages over time.}
  \label{fig:rebuildable}
\end{figure}

This research question aims to reproduce the results from Malka \emph{et al.}~\cite{malka_reproducibility_2024}, as a starting baseline.
That earlier work only built \emph{one} nixpkgs revision, the most ancient in their dataset; in our case, we rebuilt that revision alongside with 16 others, evenly spaced over time to study \emph{trends}.
\Cref{fig:rebuildable} shows the proportion of packages between 2017 and 2023 that we successfully rebuilt (not necessarily in a bitwise reproducible manner, merely ``successfully built'' for this RQ).

This proportion varies between \MinRebuildability{}\% and \MaxRebuildability{}\%, confirming previously reported findings: Nix reproducibility of build environments allows for very high rebuildability rate over time.
Note that this is not an exact replication of the revision in~\cite{malka_reproducibility_2024}, because we also included packages not explicitly listed in \texttt{release.nix}, but present in the dependency graph, whereas they did not.

\subsection{RQ1: What is the evolution of bitwise reproducible packages in nixpkgs between 2017 and 2023?}

\begin{figure}
  \includegraphics[width=0.47\textwidth, right]{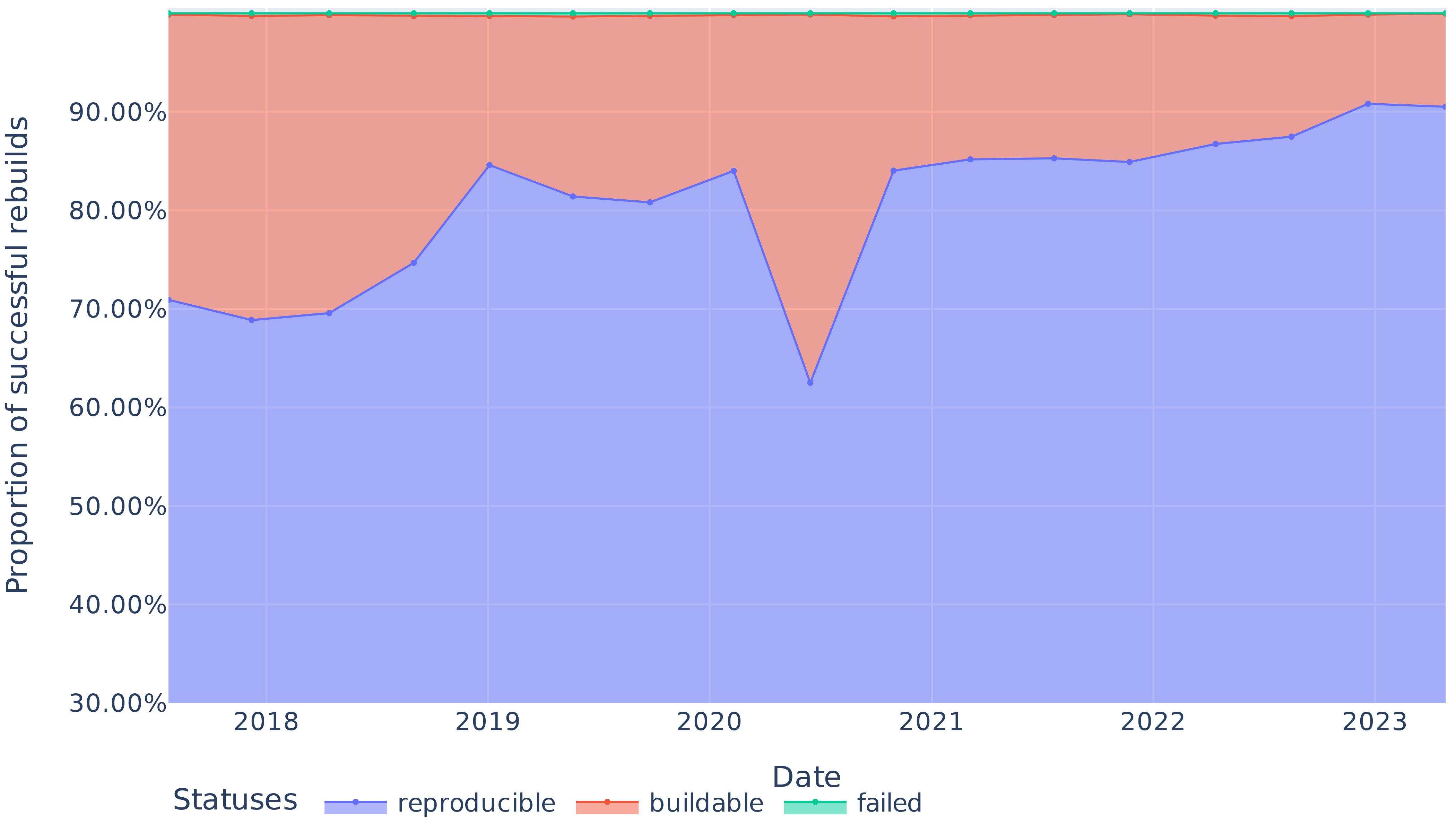}
  \caption{Proportion of reproducible, rebuildable (but unreproducible) and non-rebuildable packages over time.}
  \label{fig:repro-prop}
\end{figure}

\begin{figure}
  \includegraphics[width=0.47\textwidth, right]{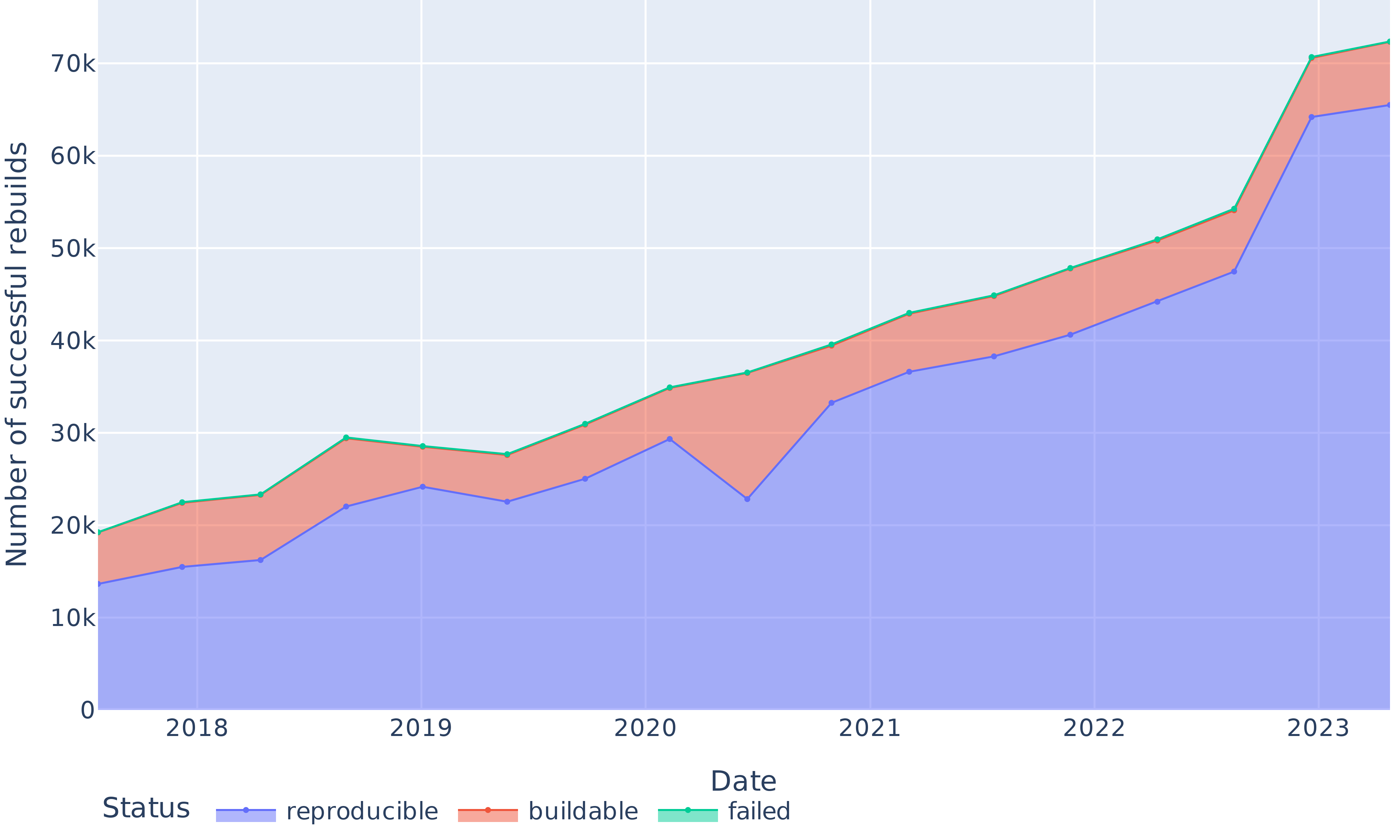}
  \caption{Absolute numbers of reproducible, rebuildable (but unreproducible) and non-rebuildable packages over time.}
  \label{fig:repro-abs}
\end{figure}

Apart from a significant regression in 2020, we obtain bitwise reproducibility levels between \MinReproducibilityPropRounded{}\% and \MaxReproducibilityPropRounded{}\% (see \Cref{fig:repro-prop}).
The trends in \Cref{fig:repro-abs} show that the absolute number of bitwise reproducible packages has consistently gone up and followed the fast growth of the package set.
The only exception is the data point for June 2020, where the number of reproducible packages dropped even though the total number of packages grew.
We study and explain this reproducibility regression in \Cref{sec:unreproducible-packages} below.

\begin{figure}
  \includegraphics[width=0.47\textwidth, right]{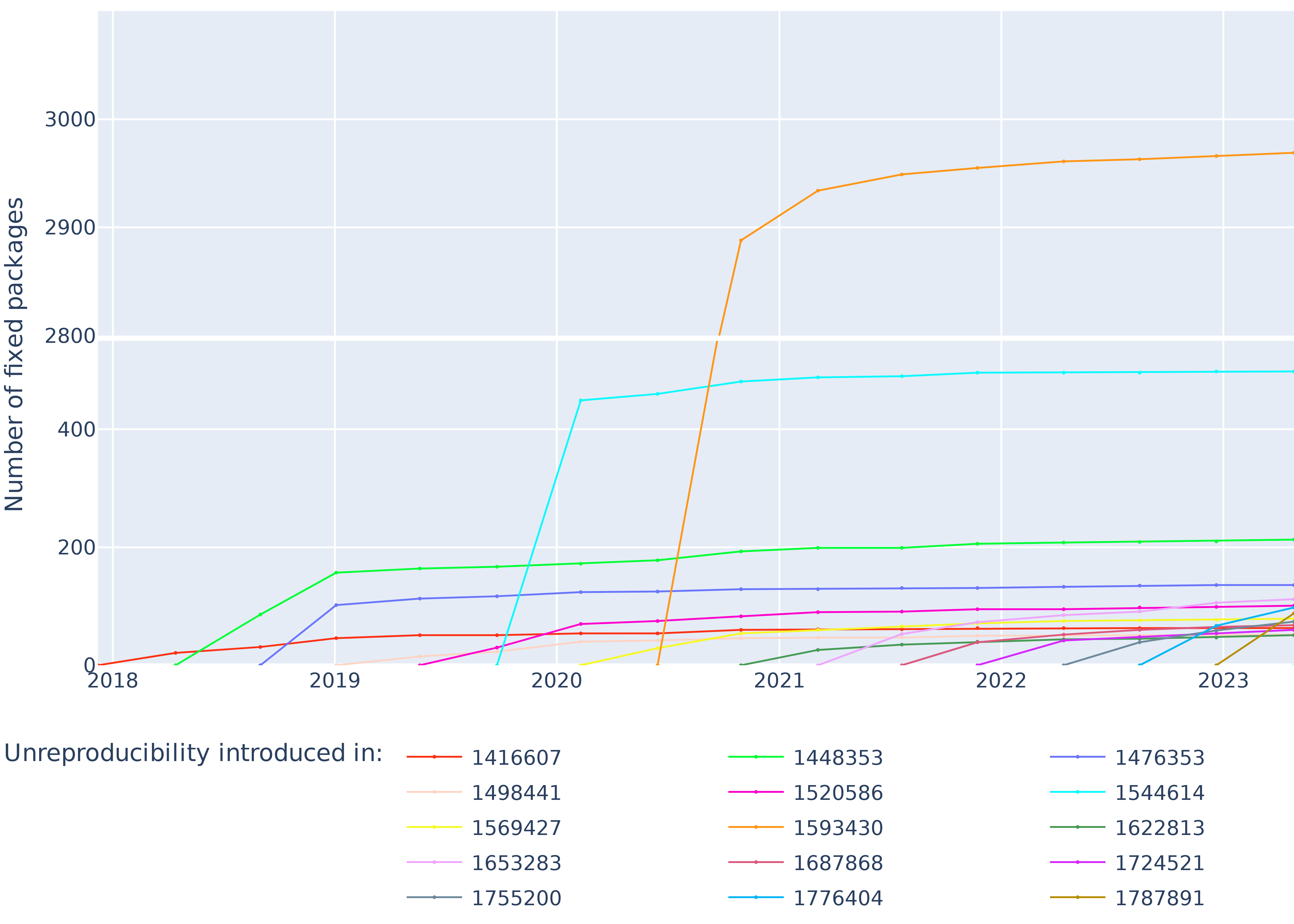}
  \caption{Evolution of the cumulative number of fixes to non-reproducibilities, separated by the revision in which the non-reproducibilities were introduced.}
  \label{fig:cumulative-fixes}
\end{figure}

\Cref{fig:cumulative-fixes} shows for each revision the cumulative amount of unreproducibilities introduced by that revision getting fixed over time.
The large slope between the two first points of each plot indicates that most of the unreproducibilities introduced in a revision are fixed in the next revision, on average \UnreproducibilityFixedNextRevision{}\% of them (even raising to \UnreproducibilityFixedNextRevisionWithRegression{}\% if we account for packages fixed after the 2020 reproducibility regression).

\begin{figure}
  \includegraphics[width=0.47\textwidth, right]{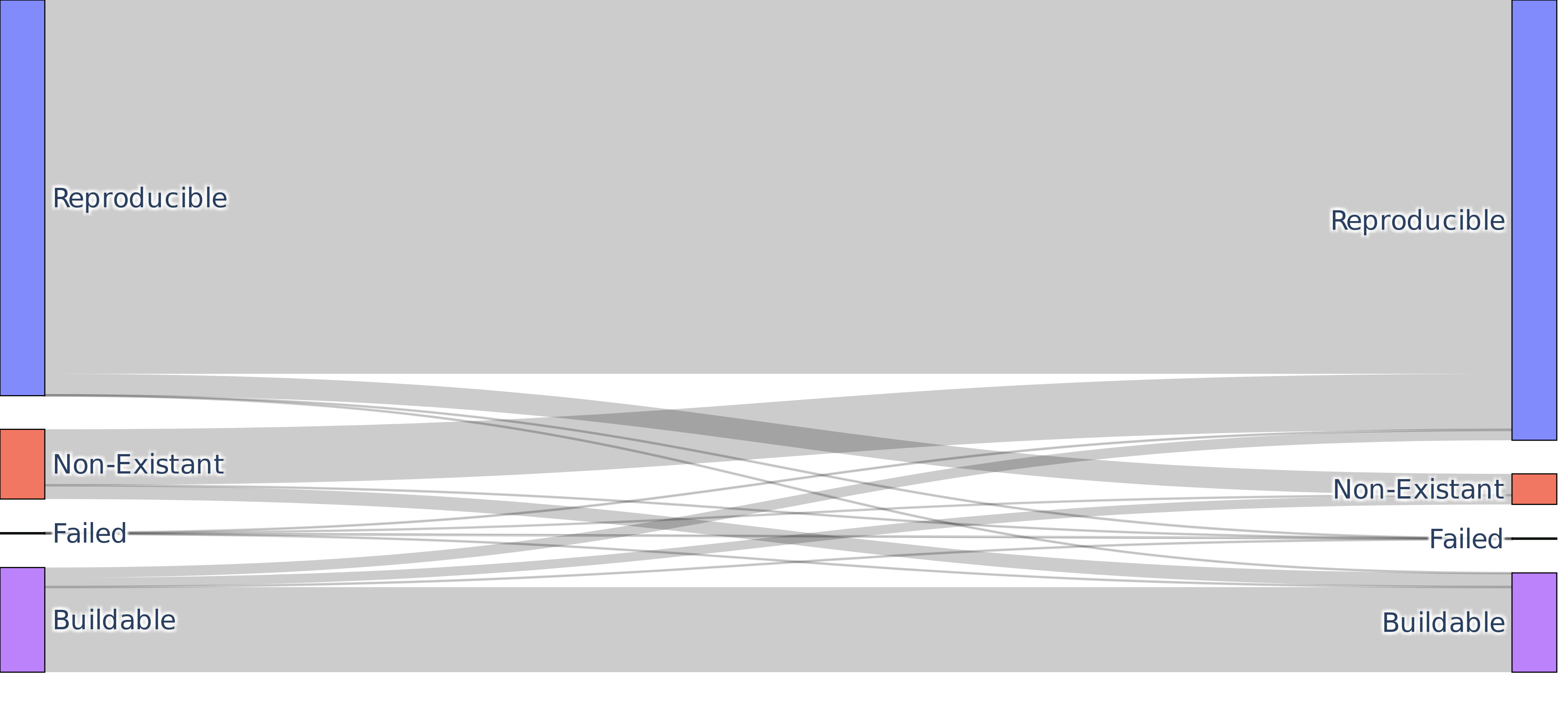}
  \caption{Sankey graph of the average flow of packages between two revisions, excluding the revision from June 2020, considered as an outlier.}
  \label{fig:average-evolution}
\end{figure}

Excluding the June 2020 revision (corresponding to the observed bitwise reproducibility regression), \Cref{fig:average-evolution} depicts the average evolution of the package set between two consecutive revisions.
In particular, only \AverageReproToBuildable{}\% of the packages transition from \textit{reproducible} to \textit{buildable but not reproducible} status between two revisions, while \AverageBuildableToRepro{}\% of the \textit{buildable but not reproducible} packages become reproducible.
The average growth rate of the package set is \AverageGrowthRate{} and, on average, \ReproducibleAddedPackageProp{}\% of the new packages are reproducible.

\subsection{RQ2: What are the unreproducible packages?}
\label{sec:unreproducible-packages}

\begin{figure}
  \includegraphics[width=0.47\textwidth, right]{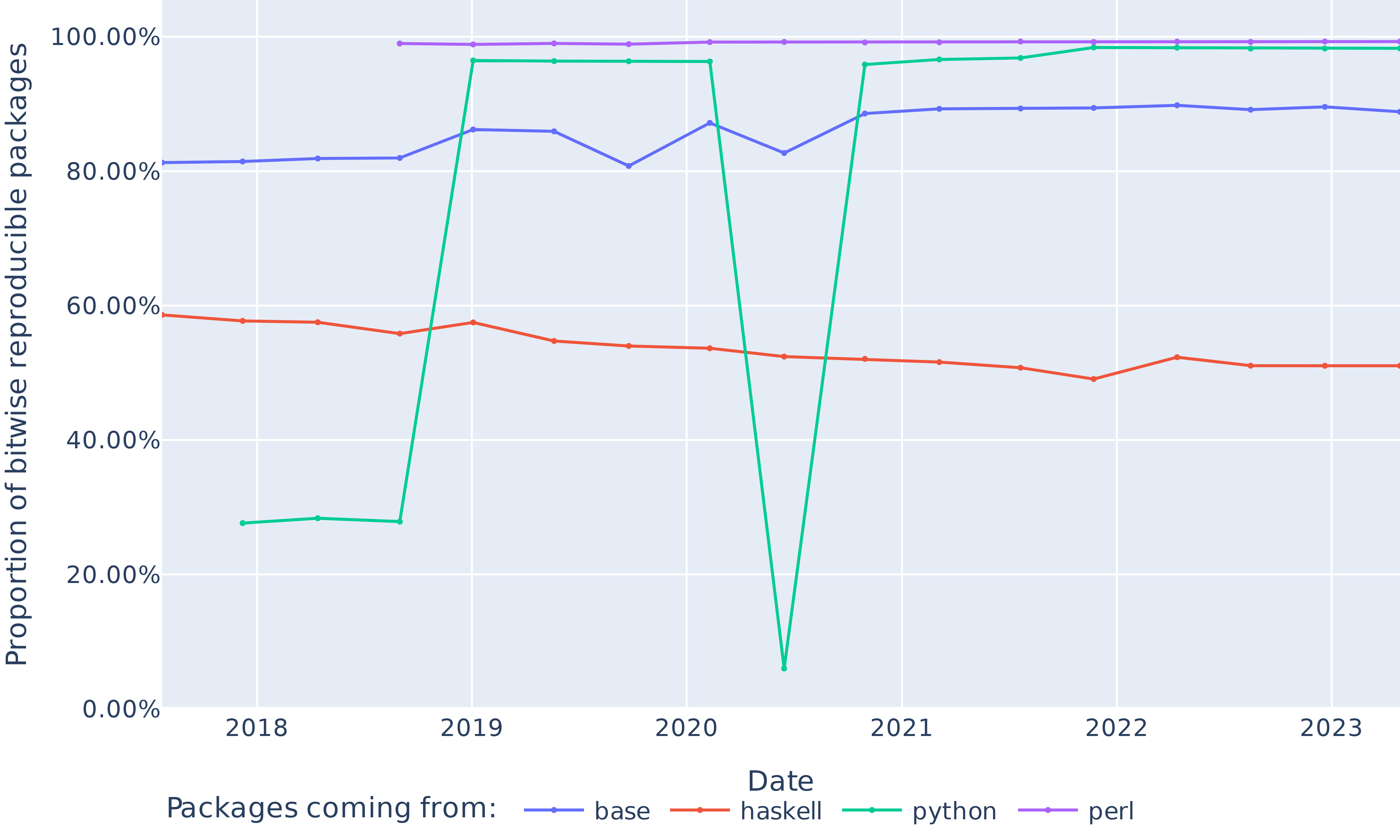}
  \caption{Proportion of reproducible packages belonging to the three most popular ecosystems and the base namespace of nixpkgs.}
  \label{fig:reproducibility-ecosystems}
\end{figure}

We observe large disparities both in trends and in reproducibility by package ecosystem (see \Cref{fig:reproducibility-ecosystems}).
In the top three most popular ecosystems, Perl has consistently maintained a proportion of reproducible packages above 98\%, while Haskell reproducibility rate stagnated around the 60\% mark, even decreasing by more than 7 percentage points during the time of our study.\footnote{A fix to a long-standing reproducibility issue has been introduced in \href{https://downloads.haskell.org/ghc/9.12.1/docs/users_guide/9.12.1-notes.html}{GHC 9.12.1} (new \texttt{-fobject-determinism} flag). As of January 2025, nixpkgs does not make use of this flag to improve Haskell build reproducibility.}
The Python ecosystem on the contrary sees a positive evolution over time, with reproducibility rates as low as 27.64\% in December 2017, reaching 98.28\% in April 2023.
As can be seen on \Cref{fig:reproducibility-ecosystems}, the Python ecosystem has however known a major dip in reproducibility in June 2020, with a drop to 6.01\% (for almost $-90$ percentage points!).

\begin{figure}
  \includegraphics[width=0.47\textwidth, right]{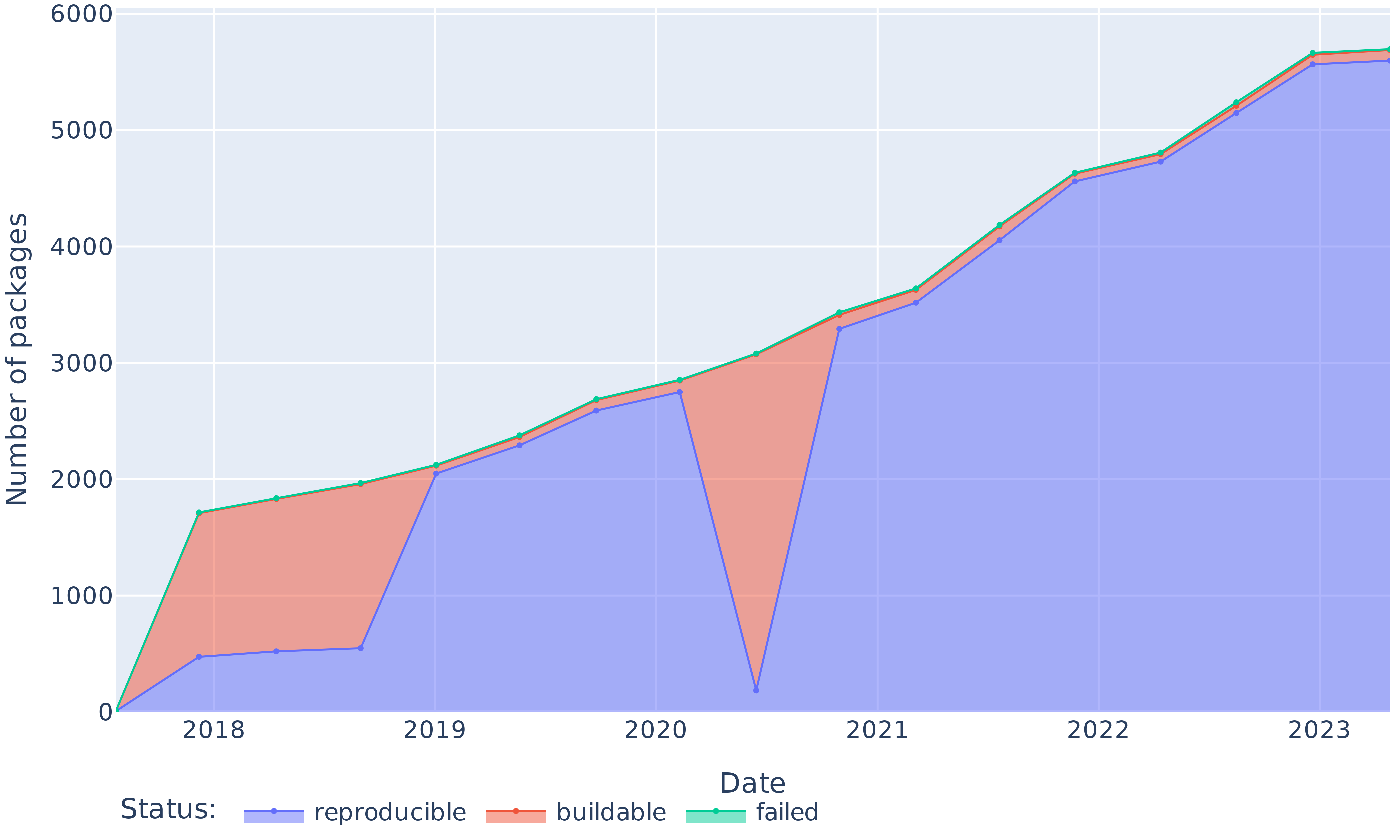}
  \caption{Evolution of the absolute number of reproducible, rebuildable (but unreproducible) and non-rebuildable packages from the Python ecosystem.}
  \label{fig:python-reproducibility-abs}
\end{figure}

As can be seen in \Cref{fig:python-reproducibility-abs}, this dip in the proportion of reproducible Python packages can be explained by a large number of packages transitioning from the reproducible to the buildable but not reproducible state, indicating a regression in bitwise reproducibility at that time.
To identify the root cause of that regression, we used a git bisection on the nixpkgs repository between June and October 2020 in order to identify the commit fixing the unreproducibility issue, and derived that the root cause of the regression was an update in the \texttt{pip} executable changing the behavior of byte-compilation.\footnote{Details can be found in \url{https://github.com/pypa/pip/issues/7808}.}

\begin{figure}
  \includegraphics[width=0.47\textwidth, right]{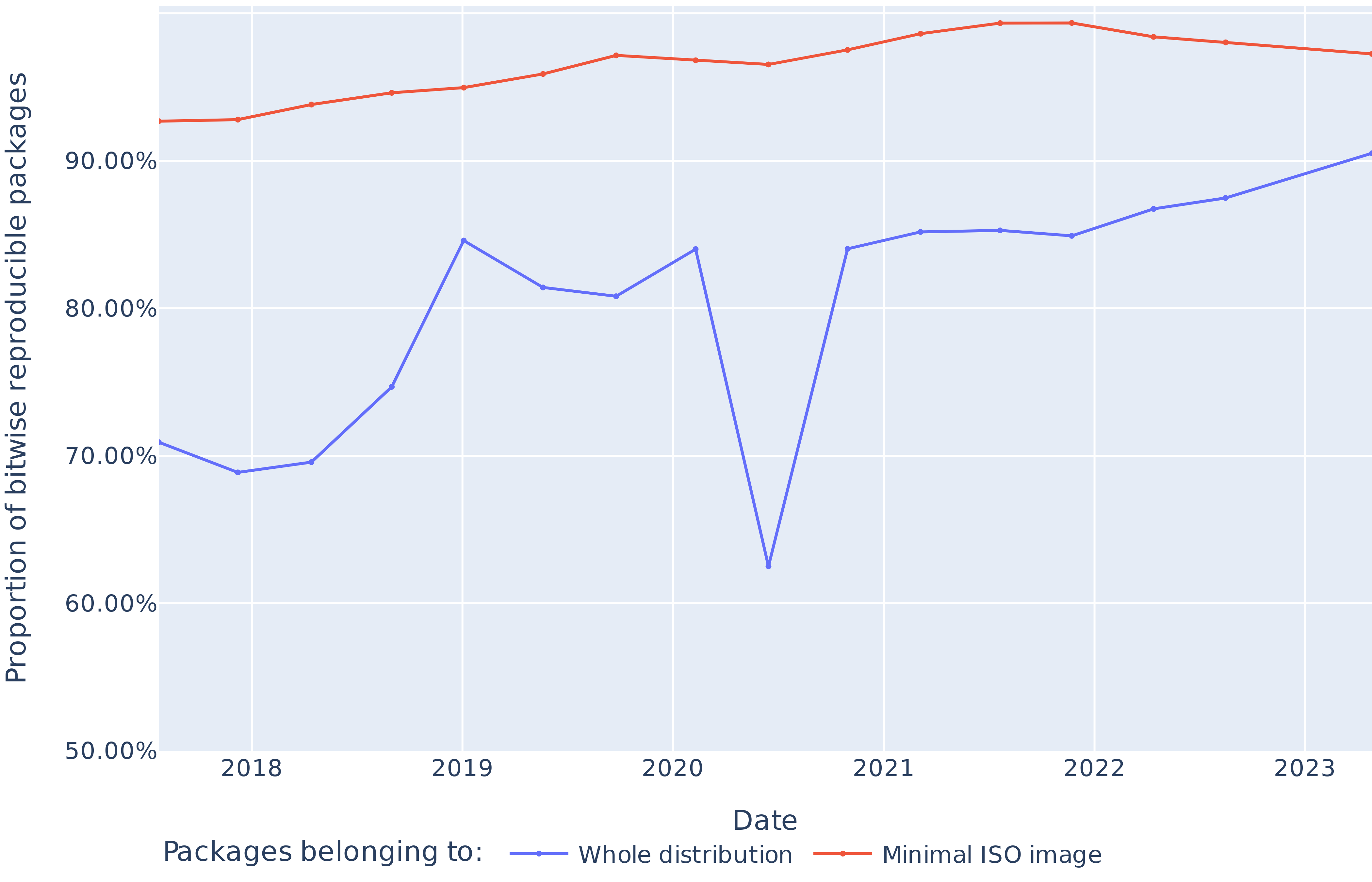}
  \caption{Evolution of the proportion of reproducible packages belonging to the minimal ISO image and in the entire package set.}
  \label{fig:minimal-iso-reproducibility}
\end{figure}

\Cref{fig:minimal-iso-reproducibility} shows the difference in reproducibility rates between the NixOS minimal ISO image---a package set for which there exists community-based reproducibility monitoring, and made of packages that are considered critical---and the whole package set.
The minimal ISO image has a very high proportion of reproducible packages in the considered period, with rates consistently higher than 95\% starting from May 2019.
Its reproducibility rates however do not follow the evolution of the overall package set, such that observing the reproducibility of the minimal ISO image does not give any clue about the reproducibility of the package set as a whole.

\subsection{RQ3: Why are packages unreproducible?}

We identified four heuristics that are both \emph{effective} (they give a large number of matches in our diffoscope dataset) and \emph{relevant} (they correspond to a software engineering practice that can be changed) to automatically determine why packages are not bitwise reproducible:
\begin{itemize}
\item \textbf{Embedded dates:} presence of build dates in various places in the built artifacts (version file, log file, file names, etc.). To detect dates embedded in build outputs, we look specifically for the year 2024 in added lines, given that we ran all our builds during 2024, but none of the historical builds ran during this year.
\item \textbf{\texttt{uname} output:} some unreproducible builds embed build information such as the machine that the build was run on. While the real host name is hidden by the Nix sandbox, other pieces of information (such as the Linux kernel version or OS version, reported by \texttt{uname}) are still available. %
\item \textbf{Environment variables:} some builds embed some or all available environment variables in their build outputs. This typically causes unreproducibility because an environment variable containing the number of cores available to build on the machine is set by Nix.
\item \textbf{Build ID:} some ecosystems (Go for example) embed a unique (but not deterministic) build ID into the artifacts.
\end{itemize}

\begin{figure}
  \includegraphics[width=0.47\textwidth, right]{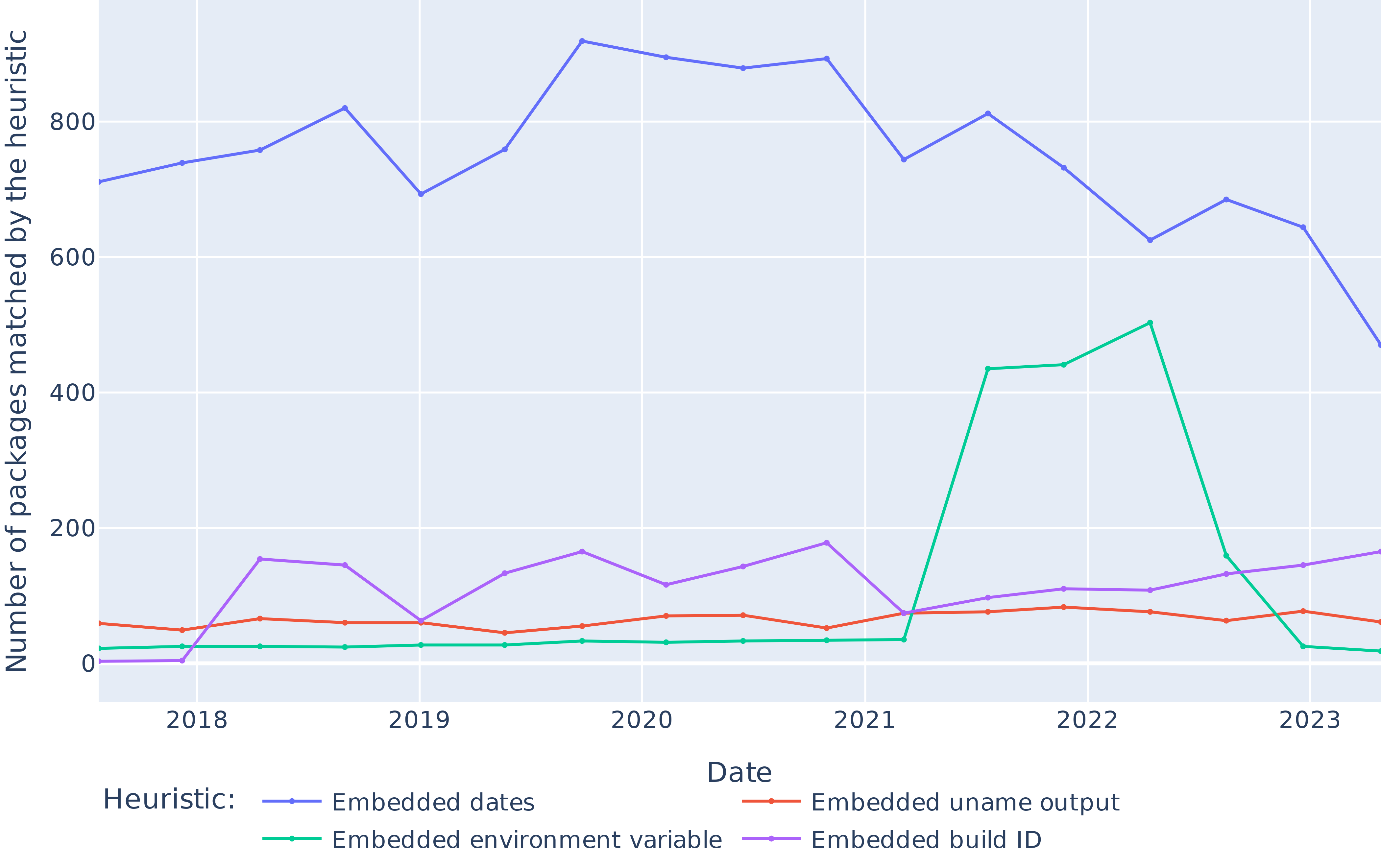}
  \caption{Evolution of the number of packages for which we generated diffoscopes that are matched by each of our heuristics, over time.}
  \label{fig:evolution-heuristics}
\end{figure}

Despite a well-known recommendation by the Reproducible Builds project to avoid embedding build dates in build artifacts to achieve bitwise reproducibility, we find that \PackagesWithEmbeddedDate of our \TotalPackagesWithDiffoscope packages with diffoscopes contain a date, accounting for \PropPackagesWithEmbeddedDate{}\% of them.
Still in the most recent nixpkgs revision, we find \PackagesWithDateLastRevision instances of this non-reproducibility cause, showing that embedding dates is still an existing practice.
Additionally, we find that embedded environment variables and build IDs each account for \PropPackagesWithEmbeddedVar{}\% of our unreproducible packages.
Finally, we find \texttt{uname} outputs in \PackagesWithEmbeddedUname of our unreproducible builds (\PropPackagesWithEmbeddedUname{}\%), \UnameMatchThatAreDates of which also include a date as part of the \texttt{uname} output.
\Cref{fig:evolution-heuristics} shows the evolution of the number of packages matched by each heuristic over time.
Altogether, a total of \PropPackagesMatchedAtLeastOnce{}\% of the packages for which we have generated a diffoscope have an unreproducibility that can be explained by at least one of our heuristics.

We evaluate the precision of each heuristic by manually verifying \num{500} matched lines for every heuristic and counting false positives.
For the \textit{uname, build ID, and environment variables} heuristics, we report a precision of 100\%.
For the date heuristic, we obtain a precision of 97.8\% (11 false positives).

\subsection{RQ4: How are unreproducibilities fixed?}

Our analysis of 100 randomly sampled reproducibility fixes showed that, in most cases, authors are not aware that they are fixing a reproducibility issue, or they do not mention it: in 93 instances, we did not find any trace of reproducibility being mentioned.
In fact, in 76 cases we found out that the reproducibility fix was just a routine package update, and it is very likely that the package becoming reproducible had more to do with changes in the upstream software than with a conscious action from the package maintainer.
Other fixes included internal nixpkgs changes that had reproducibility impact as a \textit{side effect} of the main reason behind the change.

Our study of the 15 most impactful fixes suggests that those are more often done with the intent of fixing reproducibility: in more than half of them (8 out of 15) the author mentioned and documented the reproducibility issue being fixed and 9 of them were changes internal to nixpkgs.
Some large-scale reproducibility fixes were still package updates like toolchain updates.

\section{Discussion}
\label{sec:discussion}

This work brings valuable insights to the ongoing discussions about software supply chain security, reproducible builds, and functional package management.

\paragraph*{False myth: reproducible builds (R-B) do not scale}
One strongly held belief about R-B is that they work well in limited contexts---either selected ``critical'' applications (e.g., cryptocurrencies or anonymity technology) or strictly curated distributions---but either do not scale or are not worth the effort in larger settings.
Our results can be interpreted as counter proof of this belief: in nixpkgs, the largest general-purpose repository with more than 70k packages (in \EndMonth), more than \ReproducibleLastRevision{}\% packages are nowadays bitwise reproducible from source.
R-B do scale to general purpose, regularly used software.

\paragraph*{Recommendation: invest in infrastructure for binary caches and build attestations}
To reap the security benefits of the R-B vision~\cite{lamb_reproducible_2022} in practice, users need multiple components: (1) signed attestations by each independent builder, publishing the obtained checksums at the end of a build; (2) caches of built artifacts, so that users can avoid rebuilding from source; (3) verification technology in package managers to verify that built artifacts in (2) match the consensus checksums in (1).
With few exceptions, the infrastructure to support all this does not exist yet.
Now that the ``R-B do not scale'' excuse is out of the picture, we call for investments to develop this infrastructure, preferably in a distributed architectural style to make it more secure, robust, and sustainable.

\paragraph*{False myth: Nix implies bitwise reproducibility}
Contradictory to the previous myth, there is also a belief that Nix (or functional package management more generally) \emph{implies} bitwise reproducibility.
Our results disprove this belief: about \BuildableLastRevision{}\% of nixpkgs packages are not bitwise reproducible.
This is not surprising, because R-B violations happen for multiple reasons, only some of which are mitigated by FPM \emph{per se}.

\paragraph*{Recommendation: use FPM for bitwise reproducibility and rebuildability}
Still, Nix appears to perform really well at bitwise reproducibility ``out of the box'', and even more so at rebuildability (above 99\%).
Based on this, we recommend to use Nix, or FPM more generally, for all use cases that require either mere rebuildability or full bitwise reproducibility.
At worse, they provide a very good starting point.

This work has investigated the causes of the lingering \emph{non} reproducibility in nixpkgs, but not those of reproducibility; it would be interesting to know why Nix performs so well, possibly for adoption in different contexts.
It is possible that bitwise reproducibility is an emerging property of FPM, or that it comes from technical measures during build like sandboxing, or that Nix is simply benefiting (more than other distributions?) from the decade-long efforts of the R-B project in upstream toolchains.
Exploring all this is left as future work.

\paragraph*{QA monitoring for bitwise reproducibility}
The significantly higher (and stably so) bitwise reproducibility performances of the NixOS minimal ISO image (see \Cref{fig:minimal-iso-reproducibility}) suggests that quality assurance (QA) monitoring for build reproducibility is an effective way to increase it.
The fact that Debian uses a similar approach with good results~\cite{lamb_reproducible_2022} is compatible with this interpretation.
A deeper analysis of the relationship between QA monitoring and reproducibility rate is out of scope for this work, but it is quite likely that extending reproducibility monitoring to all packages will result in an easy win for nixpkgs (assuming bitwise reproducibility is seen as a project goal).

\section{Threats to validity}
\label{sec:threats}

\subsection{Construct validity}

We rebuilt Nix packages from revisions in the period \StartMonth--\EndMonth, for about 6 years.
That is enough to observe arguably long-term trends, which also appear to be stable in our analysis; except for one temporary regression, which we analyzed and explained.
Still, we cannot exclude significant differences in reproducibility trends before/after the studied period.

Due to the high computation cost, build time, and environmental impact of rebuilding general purpose software from source, we sampled nixpkgs revisions to rebuild uniformly (every \SamplingPeriod{} months) within the studied period, totaling \BuildHours hours of build time.
We cannot exclude trend anomalies \emph{between sampled revisions} either, but that seems unlikely due to the stability of the observed long-term trends.
More importantly, this means that we are less likely to catch short-spanned reproducibility regressions, which would be introduced and fixed within the same period between two sampled revisions.
This can be improved upon by sampling and rebuilding more nixpkgs revisions, complementing this work.

When rebuilding a given package, we relied on the Nix binary cache for all its transitive dependencies.
As we have rebuilt \emph{all} packages from any sampled nixpkgs revisions, our package coverage is complete.
But this way we have not measured the impact of unreproducible packages on transitive reverse dependencies, i.e., how many \emph{additional} packages become unreproducible if systematically built from scratch, including all their dependencies?
Our experiment matches real-world use cases and is hence adequate to answer our RQs, but it would still be interesting to know.

Also, during rebuilds we have not attempted to re-download online assets, but relied on the Nix cache.
Hence, we have not measured the impact of lacking digital preservation practices on reproducibility and rebuildability.
It would be interesting and possible to measure such impact by, e.g., first trying to download assets from their original hosting places and, for source code assets, fallback to archives such as Software Heritage~\cite{dicosmo_2017_swh} in case of failure.

\subsection{External validity}

We have rebuilt packages from nixpkgs, the largest cross-ecosystem FOSS distribution, using the Nix functional package manager (FPM).
Other FPMs with significant user bases exist, such as Guix~\cite{courtes_functional_2013}.
Given the similarity in design choices, we do not have reasons to believe that Guix would perform any differently in terms of build reproducibility than Nix, but we have not empirically verified it; doing so would be useful complementary work.

Neither did we verify historical build reproducibility in classic FOSS distributions (e.g., Debian, Fedora, etc.), as out of the FPM scope.
Doing so would still be interesting, for comparison purposes.
It would be more challenging, though, due to the fact that outside the FPM context, it is significantly harder to recreate exact build environments from years ago.

\section{Conclusion}
\label{sec:conclusion}

In this work we conducted the first large-scale experiment of bitwise reproducibility in the context of the Nix functional package manager, by rebuilding \TotalBuilds packages coming from \RevisionsBuilt revisions of the nixpkgs software repository, sampled every \SamplingPeriod{} months from \StartYear to \EndYear.
Our findings show that bitwise reproducibility in nixpkgs is very high and has known an upward trend, from \MinReproducibilityPropRounded{}\% in \StartYear to \MaxReproducibilityPropRounded{}\% in \EndYear.
The mere ability to rebuild packages (whether bitwise reproducibly or not) is even higher, stably around 99.8\%.

We have highlighted disparities in reproducibility across ecosystems that coexist in nixpkgs, as well as between packages for which bitwise reproducibility is actively monitored and the others.
We have developed heuristics to understand common (un)reproducibility causes, finding that \HeuristicDate{}\% of unreproducible packages were embedding the date during the build process.
Finally, we studied reproducibility fixes and found out that only a minority of changes inducing a reproducibility fix were done intentionally; the rest appear to be incidental.

\section*{Data availability statement}

The dataset produced as part of this work is archived on and available from Zenodo~\cite{replication_metadata, replication_diffoscopes}.
A full replication package containing the code used to run the experiment describe in this paper is archived on and available from Software Heritage~\cite{replication-code:1.0}.

\section*{Acknowledgments}

This work would have not been possible without the availability of historical data from the official Nix package cache at \url{https://cache.nixos.org} with its long-standing ``no expiry'' policy.
In the context of ongoing discussions to prune the cache, we strongly emphasize its usefulness for performing empirical research on package reproducibility, like this work.
We also thank the NixOS Foundation for guaranteeing us that the revisions that we needed for this research would not be garbage-collected.

\clearpage
\printbibliography{}

\end{document}